\documentclass[a4paper,11pt]{article}
\usepackage{jinstpub}
\usepackage{lineno}
\usepackage{upgreek}
\usepackage{url}
\usepackage{adjustbox}
\usepackage{tabularx}
\usepackage[normalem]{ulem}

\usepackage{amsmath,amsfonts}
\usepackage{graphicx}
\usepackage{float}
\usepackage{comment}
\usepackage{upgreek}

\title{Intensified optical camera with Timepix4 readout}

\author[a]{Erik Hogenbirk}
\author[b,c,d]{Andrei Nomerotski,}
\author[a]{Bram Bouwens,}
\author[a]{Gabriel Diaz,}
\author[a]{Shazia Farooq,}
\author[b]{Sergei Kulkov,}
\author[a]{Erik~Maddox,}
\author[b]{Ondrej Matousek,}
\author[b,d]{Peter Svihra,}
\author[a]{Henrique Zanoli}

\affiliation[a]{Amsterdam Scientific Instruments, Science Park 106, 1098 XG, Amsterdam, The Netherlands}
\affiliation[b]{Faculty of Nuclear Sciences and Physical Engineering, Czech Technical University in Prague,
	Břehová 7, Prague, Czech Republic}
\affiliation[c]{Department of Electrical and Computer Engineering, Florida International University,
	10555 West Flagler St, Miami, U.S.A}
\affiliation[d]{Institute of Physics of the Czech Academy of Sciences,
	Na Slovance 1999/2, Prague, Czech Republic}
\emailAdd{andrei.nomerotski@cvut.cz}

\abstract{We report the first characterization results of an optical time-stamping camera based on the Timepix4 chip coupled to a fully depleted optical silicon sensor and fast image intensifier, enabling sub-nanosecond scale, time-resolved imaging for single photons. The system achieves an RMS time resolution of 0.3~ns in direct detection mode without the intensifier and from 0.6 to 1.5~ns in the single-photon regime with an intensifier for different amplitude-based signal selections. This shows that Timepix4 provides a significant improvement over previous Timepix3-based cameras in terms of timing precision, and also in pixel count and data throughput. We analyze key factors that affect performance, including sensor bias and timewalk effect, and demonstrate effective correction methods to recover high temporal accuracy. The camera’s temporal resolution, event-driven readout and high rate capability make it a scalable platform for a wide range of applications, including quantum optics, ultrafast imaging, and time-correlated photon counting experiments.}    


\arxivnumber{2509.14649}

\begin{document}
	\maketitle
	   
	\flushbottom
	
	\section{Introduction}
	\label{sec:intro}
	The ability to detect, count, and time-stamp individual optical photons is becoming increasingly important across a wide range of imaging applications. This approach enables the capture of the complete spatio-temporal and, in principle, also spectral information carried by each detected photon, providing deeper insight into the underlying physical processes. Photon-counting is already well established in X-ray imaging, where sufficient photon energy allows for direct detection and time-resolved, energy-resolved measurements. High-rate readout electronics further enable rapid accumulation of statistics, making this modality highly efficient for quantitative imaging.

In the optical domain, the development of time-stamping cameras — particularly those based on the Timepix family readout chips — has enabled similar capabilities for visible and near-infrared photons \cite{timepixcam, Nomerotski2023}. These systems offer nanosecond-scale timing resolution, opening new avenues in areas such as ion imaging \cite{Zhao2017coin,Albrechtsen2023,Sandstrm2024, Bromberger2022}, optical readout of time-projection chambers (TPC) \cite{Roberts2019}, fluorescence lifetime imaging \cite{Hirvonen2017, Sen2020, Sen2020_1}, neutron detection \cite{DAmen2021,losko2021, Yang2021, Wolfertz2024} and quantum sciences \cite{Ianzano2020, Yingwen2020, Svihra2020, Sensors2020_Nomerotski, Zhang2021, Gao2022, Zhang2022,Zhukas2021_1, Zhukas2021,Kato2022}. 
Timepix-based single photon sensitive cameras with direct registration of MCP electrons with the Timepix ASIC metal pads have been also developed and tested \cite{Vallerga2014, Fiorini2018, Bolzonella2026}.

In applications that require single-photon detection, this sensitivity is achieved by coupling the camera to an image intensifier that employs a fast (typically P47 type) scintillator. This allows detection of individual optical photons with high spatial and temporal precision. As a result, time-stamping optical cameras, such as Tpx3Cam (also known as Chronos Phoebe from ASI), have emerged as a versatile platform for investigating low-light, time-resolved, and photon-statistical phenomena in both fundamental research and applied science.
The timing resolution is one of the most important figures of merit for the time-stamping cameras. Here we describe the first tests of an intensified optical Timepix4 camera, focusing on its superior timing resolution compared to the previous version of the camera with the Timepix3 chip \cite{Nomerotski2023}.

There are several types of imaging cameras capable of single-photon detection, e.g., intensified time-stamping complementary metal-oxide-semiconductor cameras (iCMOS) \cite{jachura2015shot}, intensified or electron-multiplying charge-coupled device cameras (iCCD or EMCCD) \cite{jost1998spatial, brida2008measurement, zhang2009characterization, fickler2013real, avella2016absolute, reichert2017quality, moreau2019imaging}, cameras based on direct registration of charge generated by micro-channel plates (MCP) in Timepix chips \cite{Vallerga2014, Alozy2022} and resistive anodes (PhotonPix$^{\rm{TM}}$ camera) \cite{Karl2025}; and cameras based on single-photon avalanche diodes (SPADs) \cite{charbon2014single, perenzoni2016compact, gasparini2017supertwin, bruschini2019single, morimoto2020megapixel, wojtkiewicz2024review, jirsa2025fast, Kulkov2025}. We refer the readers to existing reviews of the subject \cite{Hadfield2009, seitz2011single, Nomerotski2019}.
	
	\section{Methods}
For the tests described here, we developed an optical camera based on the Timepix4 chip bump-bonded to a silicon sensor that is adapted for optical photons. This sensor is identical to the one used in optical cameras based on Timepix3. A housing with a C-mount flange ensures that an intensifier or lenses can be correctly mounted and focused on to the sensor.

\begin{figure}[ht]
		\centering 
		\adjustbox{valign=c}{\includegraphics[width=.56\textwidth]{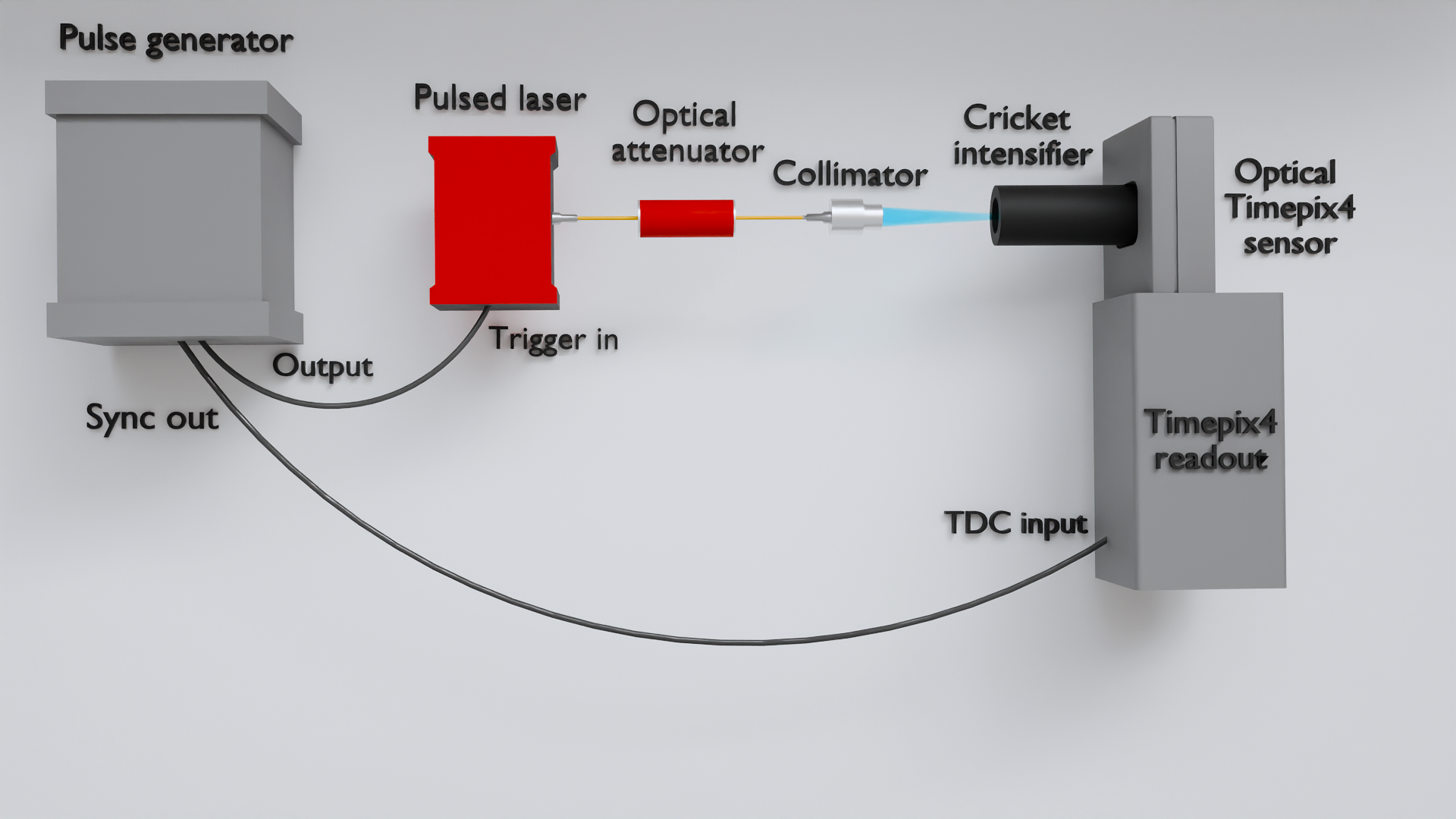}}
		\qquad
		\adjustbox{valign=c}{\includegraphics[width=.37\textwidth]{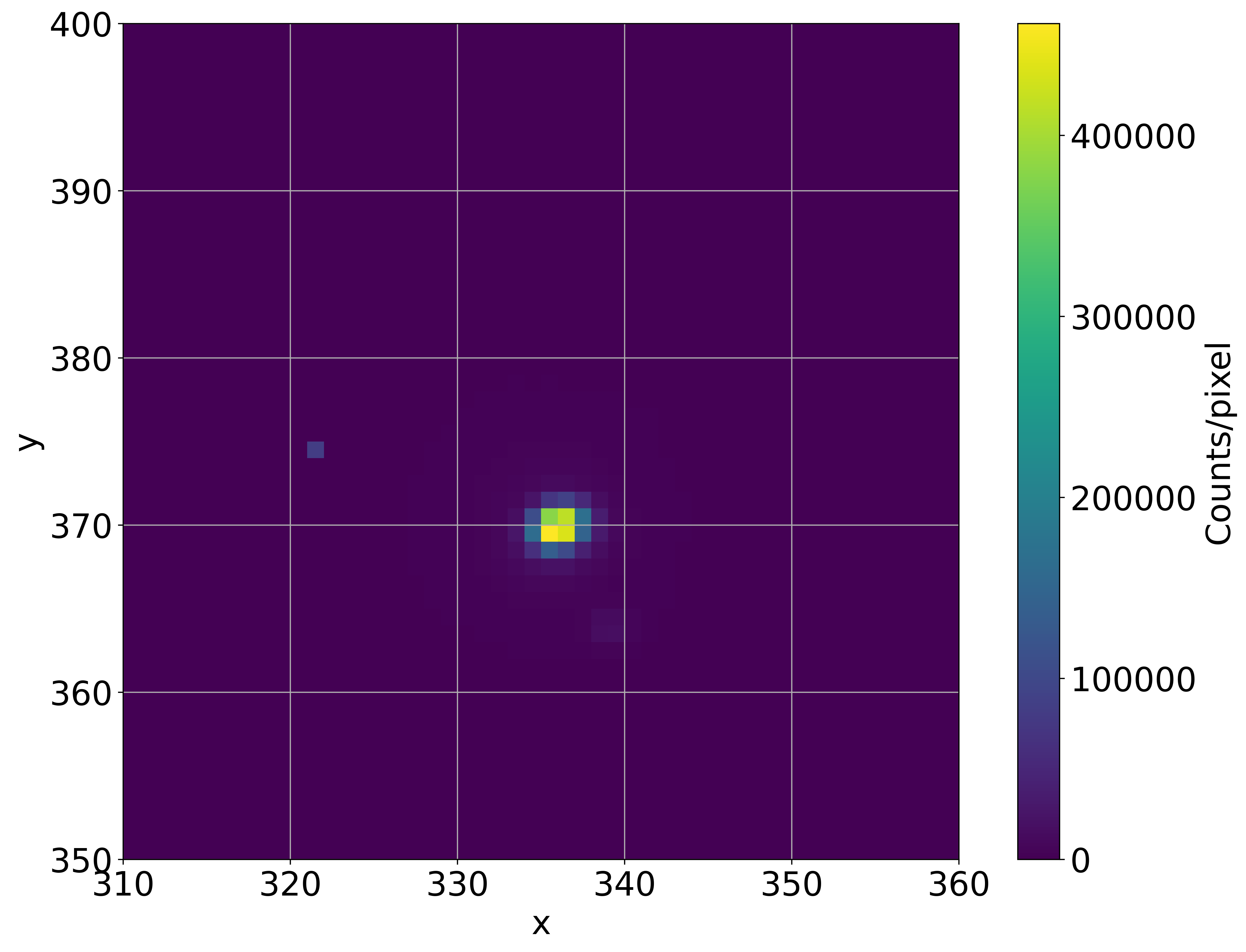}}
		\caption{\label{fig:setup} Left: Experimental setup with intensified camera and 450~nm laser  with pulse duration of 90~ps, triggered with a pulse generator running at 185~kHz. The fiber-coupled light flash had provisions to be attenuated before collimation onto the camera. The measurements have been performed with and without an intensifier, but only the intensified configuration is shown here. Right: the camera (x,y) occupancy map in configuration with an intensifier for a focused beam.}
	\end{figure}

The testing was performed with a pulsed blue 450~nm laser that illuminated a small area of the optical Timepix4 camera field of view. This was done both for intensified and non-intensified camera configurations. The timing studies relied on measurements of a delay between the laser pulse and the time-stamped pixel response. The experimental setup is schematically shown in the left part of Figure \ref{fig:setup}, and we describe it and its essential components below.

	\subsection{Timepix4 readout chip}
	
The Timepix4 is a next-generation pixel readout chip developed by the Medipix4 collaboration, designed for high-resolution, time-resolved particle and X-ray detection \cite{Llopart2022}. It features a matrix of 512 × 448 pixels with a 55~$\upmu$m × 55~$\upmu$m pixel pitch, and offers simultaneous measurement of time-of-arrival (ToA) and time-over-threshold (ToT) per pixel. The chip supports data-driven readout and multi-hit capability, enabling each pixel to independently report events without the need for a global trigger, which is critical for sparse, high-rate environments \cite{Heijhoff2022}. Its back-end architecture supports hit rates exceeding 180~Mhit/cm²/s, making Timepix4 well-suited for high-throughput, time-resolved imaging. Timepix4 measures time with a granularity of 195~ps corresponding to an RMS resolution of about 60~ps. For comparison, the Timepix3 time bin size is 1.56~ns, corresponding to a time resolution limit of 450~ps.
   
	\subsection{Optical Timepix cameras}

Optical cameras based on the Timepix3 chip have become a powerful tool for time-resolved photon detection in the visible and near-infrared domains used in a multitude of applications. The Timepix3 chip features a data-driven readout architecture, enabling each of its 256 × 256 pixels (with 55 µm pitch) to independently record both the time-of-arrival (ToA) and time-over-threshold (ToT) of incoming signals with a timing resolution of 1.56~ns and some energy discrimination via pulse-height ToT information \cite{Poikela2014}. 

In optical imaging applications, these cameras are typically coupled to optical sensors with high quantum efficiency in the 400-950~nm wavelength range \cite{Nomerotski2017}. Threshold for time-stamping a fast light flash is about 1000 photons per pixel \cite{Nomerotski2023}. The optical data-driven time-stamping concept was originally developed in 2015 for ion imaging in the velocity map imaging (VMI) applications when the first optical sensors compatible with the Timepix chip had been developed and produced \cite{timepixcam}. For single-photon applications, image intensifiers are employed, which convert single photons into fast light flashes that can be detected by the optical sensor with high efficiency. This configuration enables single-photon sensitivity, nanosecond timing, and spatial resolution within the same platform. Since the intensifier and camera are completely mechanically decoupled, the same camera can be used with different intensifiers and detection schemes, which makes this configuration particularly flexible. 

The first single-photon-sensitive Timepix camera was used for lifetime imaging in 2017 \cite{Hirvonen2017}, and the intensified Timepix3 camera was first tested for quantum applications in 2019 \cite{Ianzano2020}. At present,  optical Timepix3 cameras are increasingly used in fields such as quantum optics, ion and electron imaging, fluorescence lifetime imaging, and time-resolved spectroscopy, where precise spatio-temporal characterization of low-light signals is essential. Their ability to operate in event-driven mode without the need for external triggering makes them particularly well-suited for detecting sparse, asynchronous photon events with minimal dead time.

In the optical Timepix4 camera used for these experiments, a Timepix3 optical sensor, which is smaller in area than the full Timepix4 chip, was bump-bonded to Timepix4 and read out with SPIDR4 electronics. The leakage current of the sensor was larger than usual, which is likely caused by the edges of the sensor touching parts of the Timepix4 solder bump array. We therefore kept the bias voltage limited to 70~V in the experiments, while typically it can be safely increased to 100~V.
The camera had mounting provisions for the image intensifier packaged as Photonis Cricket$^{\rm{TM}}$, which integrates the intensifier, power supply, and back-end optics between the intensifier output and optical sensor.
The camera also implements up to four time-to-digital-converters (TDC) employing some of the peripheral Timepix4 pixels for precise time-stamping of external signals. We employed it to time-stamp the pulse generator synchronization signal used to trigger the 90~ps laser flash.
    
	\subsection{Optical setup}

The 450~nm laser (Thorlabs GSL45A) was triggered with a pulse generator at 185~kHz. The laser pulse had the minimal allowed pulse duration of 90~ps (FWHM). The produced light flash had provisions for attenuation before collimation onto the camera. In practice, the attenuation was used only in configuration with the intensifier. The right half of Figure \ref{fig:setup} shows the camera (x,y) occupancy map with the intensifier for a well-focused beam.

    \section{Results}

The primary goal of this study was to evaluate the achievable time resolution for the Timepix4 camera in optical configurations. The key question here would be to quantify the improvement of temporal resolution due to the improved time binning of the readout chip, from 1.56~ns in Timepix3 to 195~ps in Timepix4, so an improvement by a factor of eight. However, it is not clear at all if this improvement can be realized as the signal amplification in the intensifier involves light emission of the P47 scintillator, which may limit the achievable resolution improvement due to its relatively slow time response with risetime of 7~ns \cite{Winter2014}. The study below is the first attempt to quantify this effect.
	
	\subsection{Non-intensified camera}

 First we performed tests of the optical Timepix4 camera without the intensifier. In this configuration the camera is not single-photon sensitive, but detection can be achieved by a sufficiently fast and bright flash. After adjusting the focusing a small spot corresponding to several pixels was illuminated on the sensor. We selected a single pixel in this spot, and all results presented below were obtained for this single pixel, so we did not need to account for the pixel-to-pixel variability issues. We assumed and relied on the previous experience that the pixel behaviour is uniform enough across the chip.

 The blue light with the wavelength of 450~nm is absorbed near the surface in silicon with an absorption depth of about 0.5 micron. The produced carriers (holes as the sensor is p-on-n type) need to drift through the full thickness of the depleted sensor to the Timepix4 pixels. The drift time would depend on the applied bias voltage and possibly on the properties of the surface, which may require additional bias voltage to be fully depleted. Diffusion of the drifting holes may affect the time resolution at 100~ps level, but it is not expected to contribute significantly to the time resolution. 
 
\begin{figure}[ht]
		\centering 
		\adjustbox{valign=c}{\includegraphics[width=.47\textwidth]{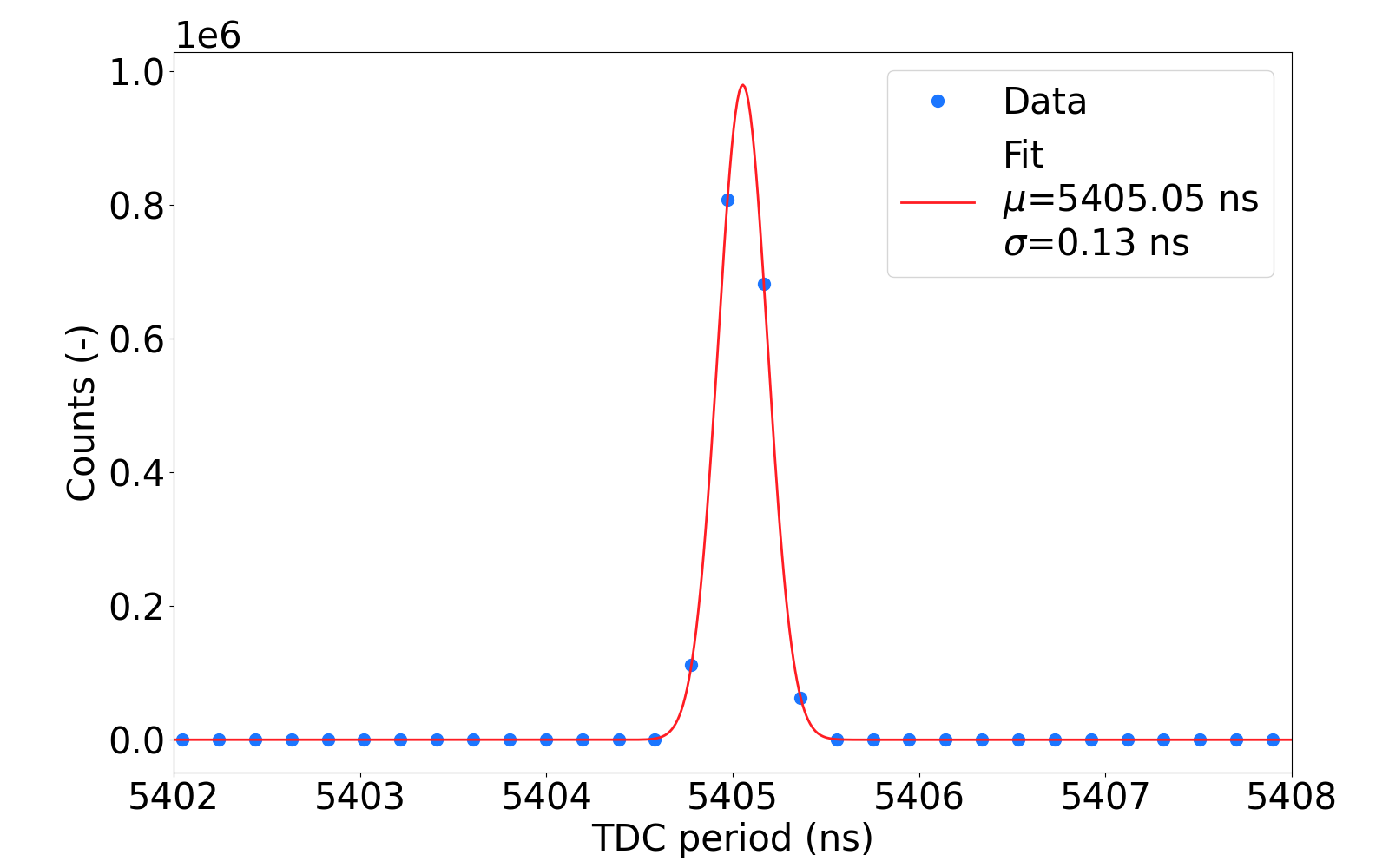}}
		\qquad
		\adjustbox{valign=c}{\includegraphics[width=.47\textwidth]{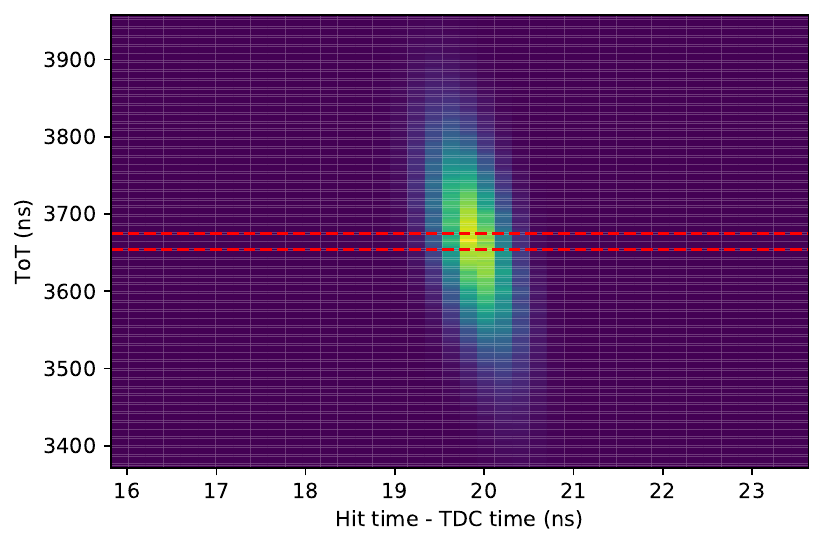}}
		\caption{\label{fig:nointensifier} Left: Distribution of measured distances between consecutive TDC pulses demonstrating the laser stability; Right: Two-dimensional distribution of time differences between the laser synchronization signal and ToA of one of the hit Timepix4 pixels versus the pixel ToT. The pixels were activated with a direct flash of light from the laser.}
	\end{figure}

 The left part of Figure \ref{fig:nointensifier} shows the distribution of measured distances between consecutive TDC synchronization pulses fit with a Gaussian function. The temporal jitter of a single laser pulse can be estimated as 0.13~ns$/\sqrt{2}=0.09$~ns, demonstrating excellent stability of the laser reference signal. The measured period of 5405~ns agrees with the laser pulsing frequency of 185 kHz.
 The right part of Figure \ref{fig:nointensifier} shows the two-dimensional distribution of time differences between the laser synchronization signal and ToA of one of the hit Timepix4 pixels activated with a flash of light from the laser versus the pixel ToT. We see that the ToT of the used pixel is contained in the 3500-3800~ns range and so does not vary much. We note that this value is due to direct illumination of the pixel during a single 90~ps laser flash. 


\begin{figure}[ht]
		\centering 
		\adjustbox{valign=c}{\includegraphics[width=.47\textwidth]{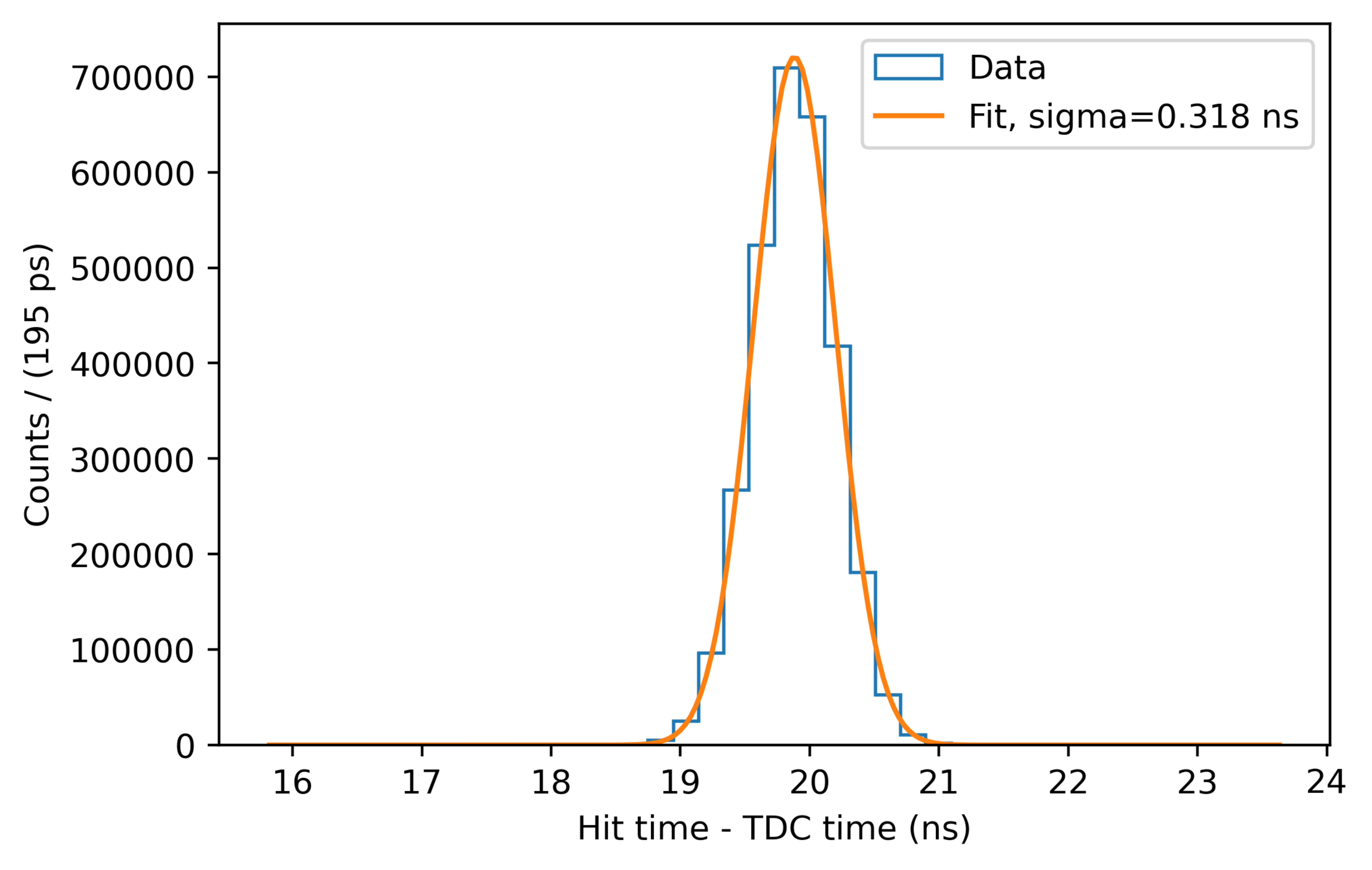}}
		\qquad
		\adjustbox{valign=c}{\includegraphics[width=.47\textwidth]{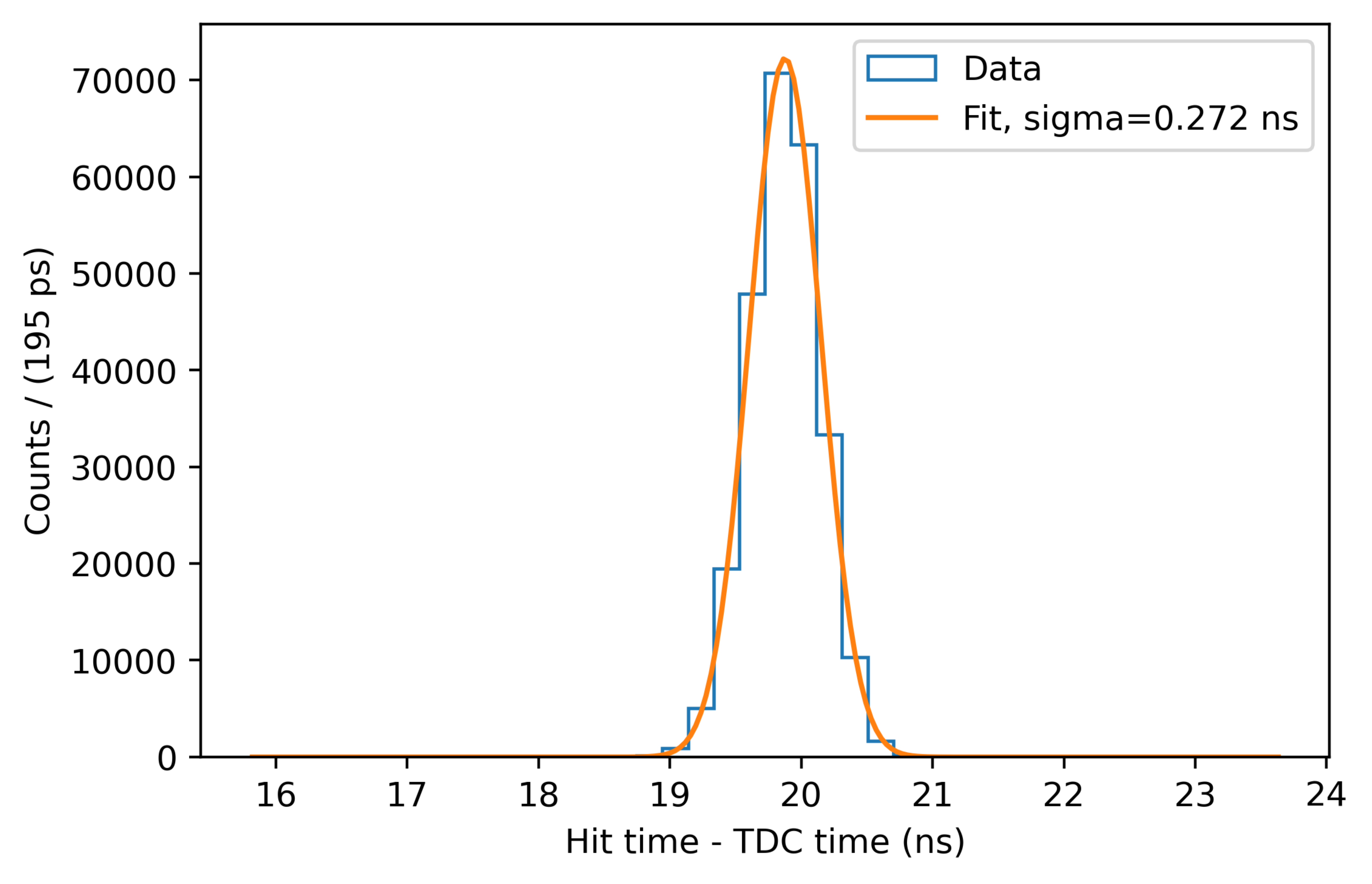}}
		\caption{\label{fig:nointensifier_1} Left: Distribution of time differences between the laser synchronization signal and ToA of one of the hit Timepix4 pixels without any TOT range selection. Right: Distribution of time differences between the laser synchronization signal and ToA of one of the hit Timepix4 pixels with TOT selection in a 25~ns slice shown in Figure \ref{fig:nointensifier} (right). The measurements were performed at a bias voltage of 50~V.}
	\end{figure}

The left part of Figure \ref{fig:nointensifier_1} shows the time difference distribution for a pixel activated with a flash of light from the laser without any TOT selection, while the right graph shows the same but with TOT selection in a 25~ns slice indicated with red dashed lines in Figure \ref{fig:nointensifier}. The time resolution (RMS) was estimated by fitting the Gaussian function to the time difference distribution. The resolution was equal to 318~ps for the whole dataset and to 272~ps for a selected 25~ns slice of ToT values. The measurements were performed at a bias voltage of 50~V.

    \subsection{Intensified camera}

We used the intensifier from Photonis with a hi-QE-blue photocathode, chevron (double-layer) MCP at its maximum gain of about $10^6$, and P47 scintillator. The intensifier was used in the Cricket$^{\rm{TM}}$ configuration with the laser flash directed to the intensifier photocathode. In this case, the laser pulse had to be strongly attenuated to provide a reasonable rate of single photons. The typically activated area of several pixels is shown in the right part of Figure \ref{fig:setup}.

\begin{figure}[ht]
		\centering 
		\adjustbox{valign=c}{\includegraphics[width=.47\textwidth]{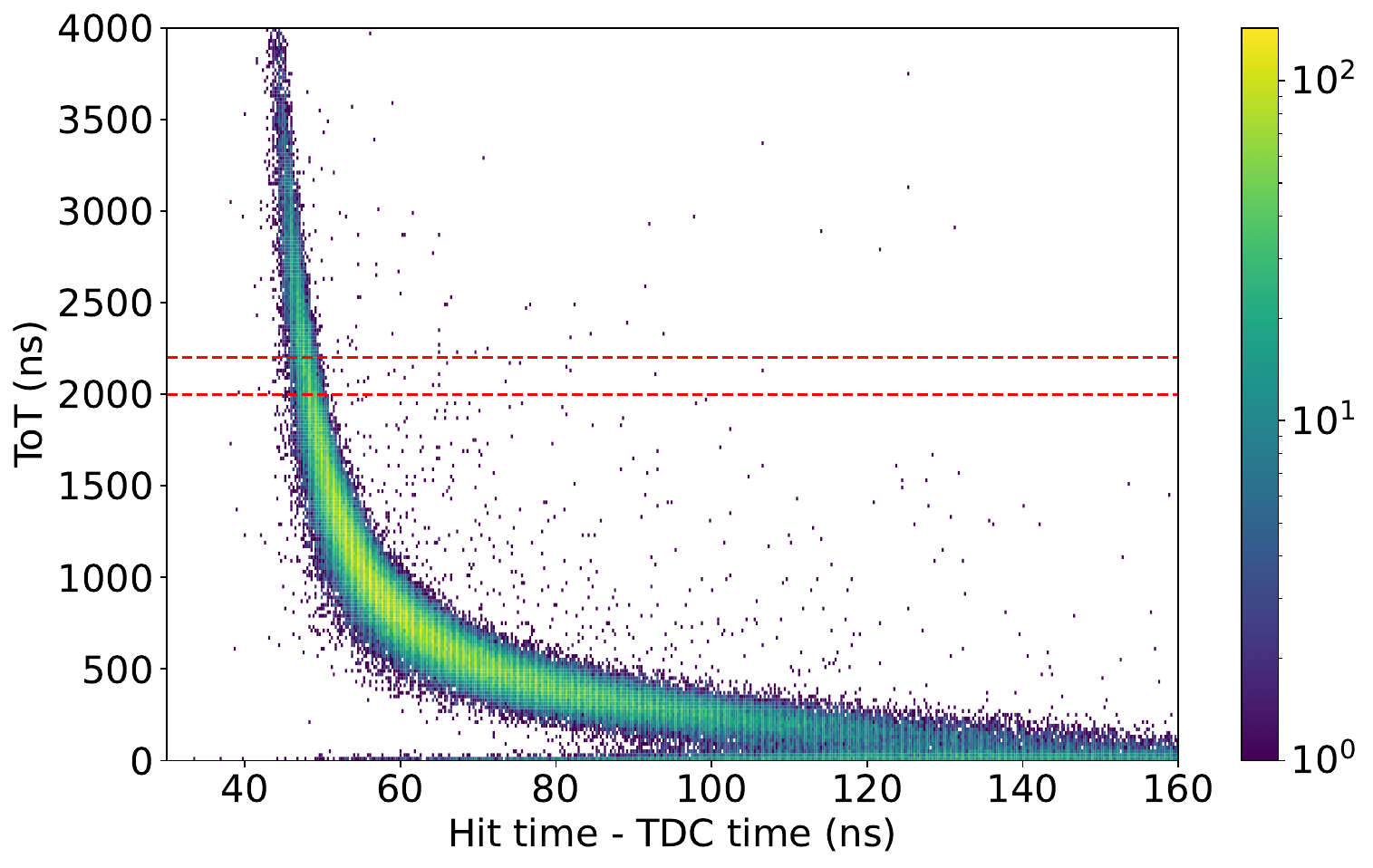}}
		\qquad
		\adjustbox{valign=c}{\includegraphics[width=.47\textwidth]{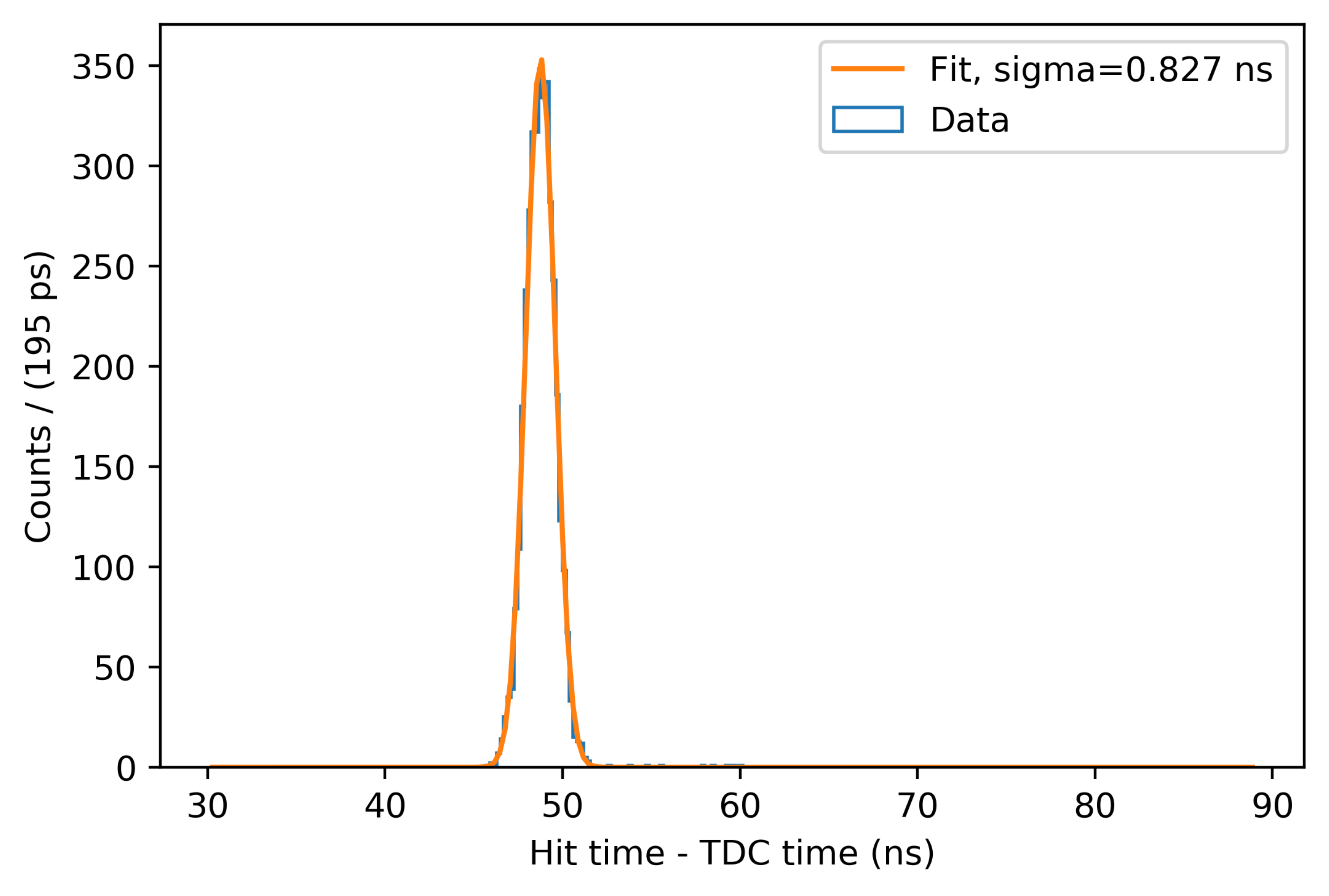}}
		\caption{\label{fig:intensifier_1} Left: Distribution of ToT versus time difference between the laser and pixel. The TOT selection in a 200~ns range, from 2000 to 2200~ns, is indicated with red dashed lines. Right: Distribution of time differences between the laser synchronization signal and ToA of one of the hit Timepix4 pixels with TOT selection in the 200~ns range fit with a Gaussian function. The measurements were performed at a bias voltage of 70~V.}
	\end{figure}

The P47 scintillator emits light over a spectrum with a maximum at 430~nm and an approximate range of 390-490~nm \cite{Winter2014}, and, thus, it behaves similarly, in terms of photon absorption, to the used 450~nm laser with photon conversions near the silicon sensor surface and considerable drift time through the full sensor depth.
The left part of Figure \ref{fig:intensifier_1} shows the two-dimensional distribution of time differences between the laser synchronization signal and ToA of one of the Timepix4 pixels versus ToT. The measurements were performed at a bias voltage of 70~V.
The distribution clearly shows the expected dependence of the delay on ToT due to the timewalk effect, addressed in detail in the following section. As an example of time resolution value we estimate it for ToT selection in the range of 2000-2200~ns, as shown in the right part of Figure \ref{fig:intensifier_1}, to be equal to 0.83~ns. This demonstrates that we can achieve sub-nanosecond resolution for the intensified (and so single-photon sensitive) optical Timepix4 camera if the intensifier gain is large enough, see further discussion in Section \ref{sec:resolution}.

        \subsection{Timewalk correction}

The timewalk terminology is used to describe the dependence of the timing response on the signal amplitude. For linear discriminators such as used in Timepix4 pixels with larger signals cross the threshold earlier than signals with smaller amplitudes. This effect can be studied and corrected for as previously demonstrated with both Timepix3 and Timepix4 chips \cite{Tsigaridas2019,Bolzonella2024, Zhao2017coin}.

\begin{figure}[ht]
		\centering 
		\adjustbox{valign=c}{\includegraphics[width=.47\textwidth]{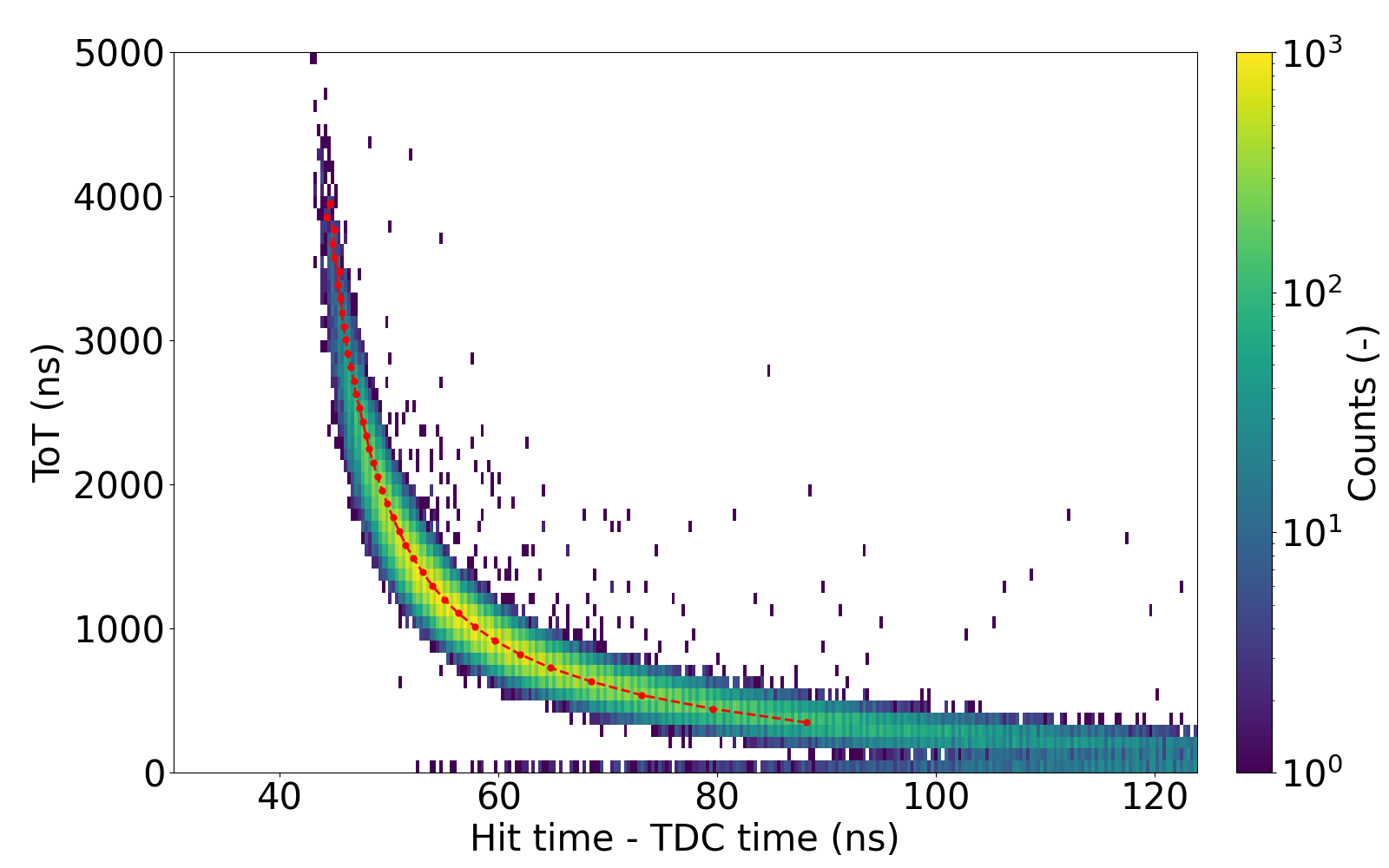}}
		\qquad
		\adjustbox{valign=c}{\includegraphics[width=.47\textwidth]{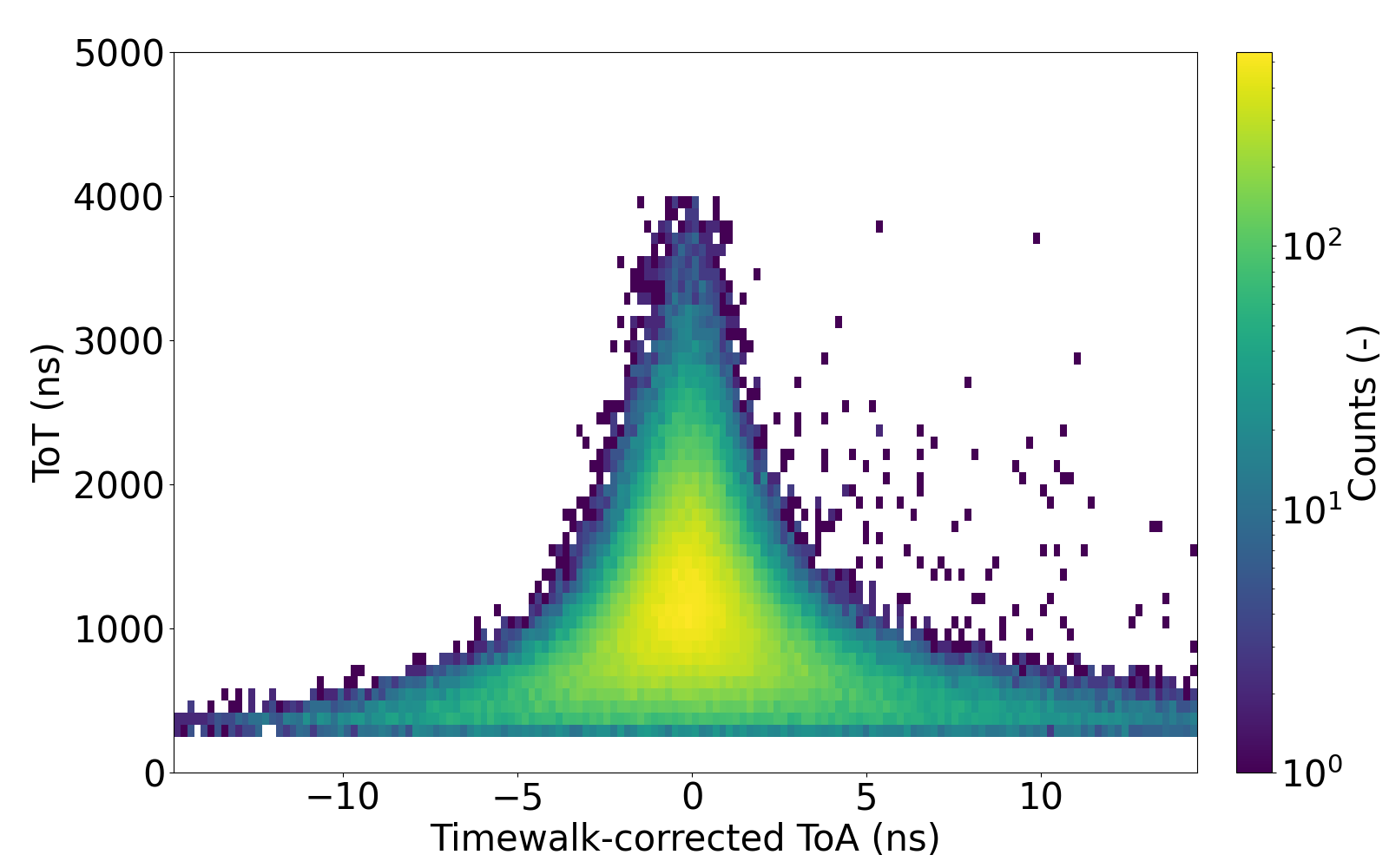}}
		\caption{\label{fig:timewalk} Left: results of the Gaussian fit for the mean values as a function of ToT. The mean values are marked with red dots on the top of the time difference versus the ToT distribution. Right: time difference distribution versus ToT after the timewalk correction.}
	\end{figure}

We performed timewalk correction by fitting the time difference distributions with a Gaussian function in 92.5~ns slices of ToT. Figure \ref{fig:timewalk} shows the results of the fit for the mean values as a function of ToT in the left part of the figure. The mean values are marked with red dots on the top of the time difference versus the ToT distribution. The right part of the figure shows the two-dimensional distribution of time difference versus ToT after the timewalk correction. The measurements were performed at a bias voltage of 70~V. One can see that the timing jitter is larger for smaller values of ToT.

\subsection{Time resolution}
\label{sec:resolution}

	\begin{figure}[ht]
		\centering 
		\adjustbox{valign=c}{\includegraphics[width=.47\textwidth]{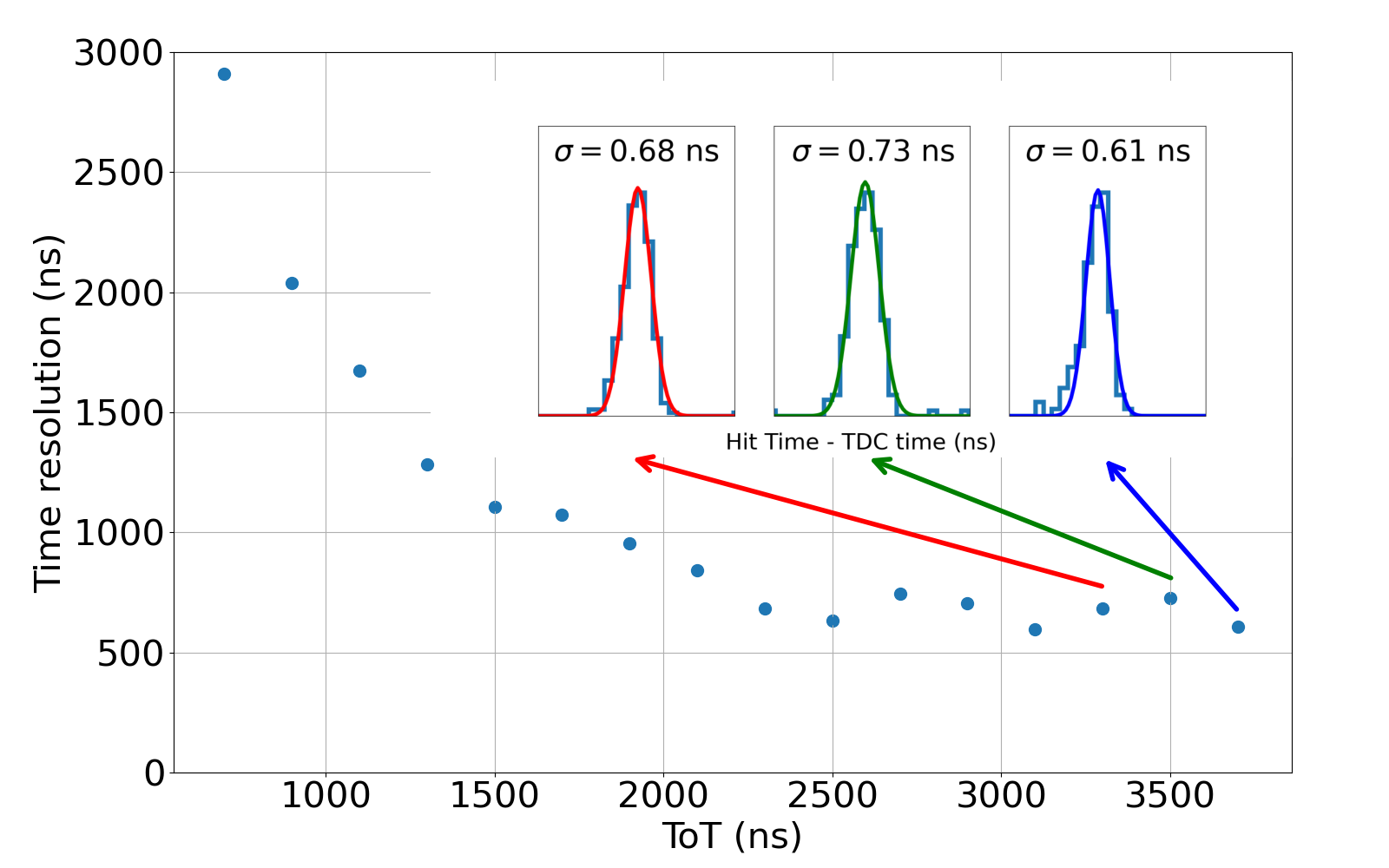}}
		\qquad
        \adjustbox{valign=c}{\includegraphics[width=.47\textwidth]{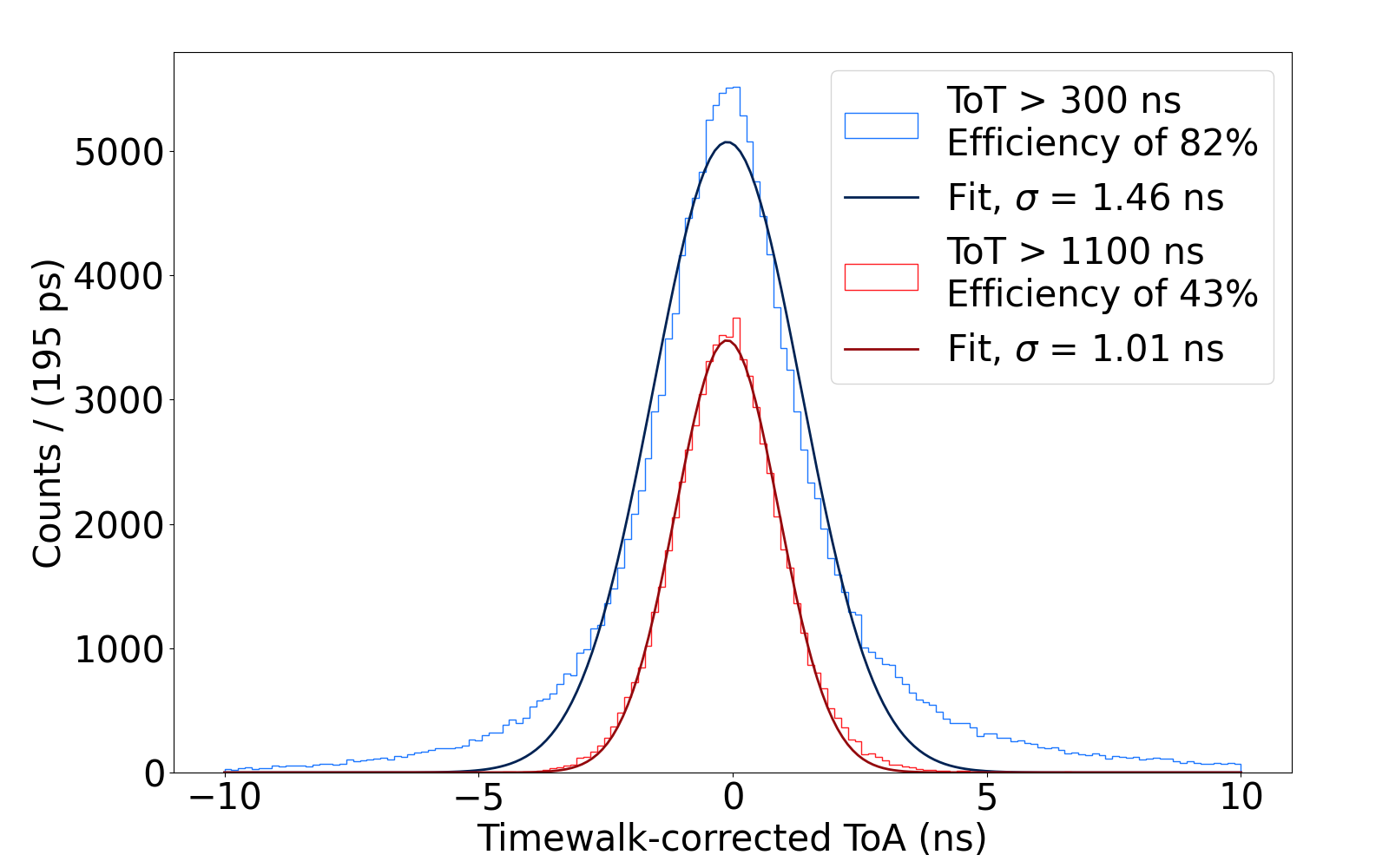}}
		\caption{\label{fig:resolution} Left: Results of the Gaussian fit for the sigma values as a function of ToT. The distribution of time differences is explicitly shown in inserts for three points with the highest ToT values. Right: Two time difference distributions after the timewalk correction, selecting ToT larger than, respectively, 300~ns and 1100~ns, fit with Gaussian functions.}
	\end{figure}

Evaluation of the timewalk correction requires performing Gaussian fits of the ToT slices. The left part of Figure \ref{fig:resolution} shows the fit results for the sigma values as a function of ToT. Distributions of time differences are
explicitly shown in inserts for the last three points with the highest ToT values. We note that the timing resolution for these ToT values is around 600-700 ps, which is very good indeed, also meaning that other contributing factors to the resolution, such as surface charge collection, diffusion in silicon and scintillator response, are under control and below this value.

After the timewalk correction, the entire time difference distribution was fit with a Gaussian function, selecting ToT larger than 300~ns, which accounted for 83\% of events. The fit is shown in the right part of Figure \ref{fig:resolution}, resulting in time resolution of 1.46~ns. The same procedure with 1100~ns ToT selection yields the resolution of 1.01~ns and efficiency of 43\%. For this study, we used the cluster based information, where ToA of the brightest pixel in the cluster is used as an estimate of the single photon time-stamp. The cluster is a collection of adjacent pixels, defined during data post-processing, which effectively represents the detector response to a single photon \cite{Nomerotski2023}.

The ToT distribution is covering a wide range from small to large ToT values due to substantial gain variations of the MCP response to photoelectrons, which is characteristic of any intensifier. The largest ToT values in the intensified case, presented in this section, reach the ToT range of the non-intensified case and, therefore, the time resolution of 600-700~ps achieved there can be directly compared to 272~ps in the non-intensified case in the right part of Figure \ref{fig:nointensifier_1}. It is likely that the difference in the resolution between these two cases can be explained primarily by the P47 scintillator contribution; we reckon that other possible contributions from the photocathode-to-MCP transit and in MCP itself are considerably smaller. We consider this as a viable confirmation that there are no fundamental showstoppers for the intensified optical Timepix4 cameras to deliver the time resolution around 600~ps if the optical gain is high enough.

    \subsection{Dependence on bias voltage}

\begin{figure}[ht]
		\centering 
		\adjustbox{valign=c}{\includegraphics[width=.47\textwidth]{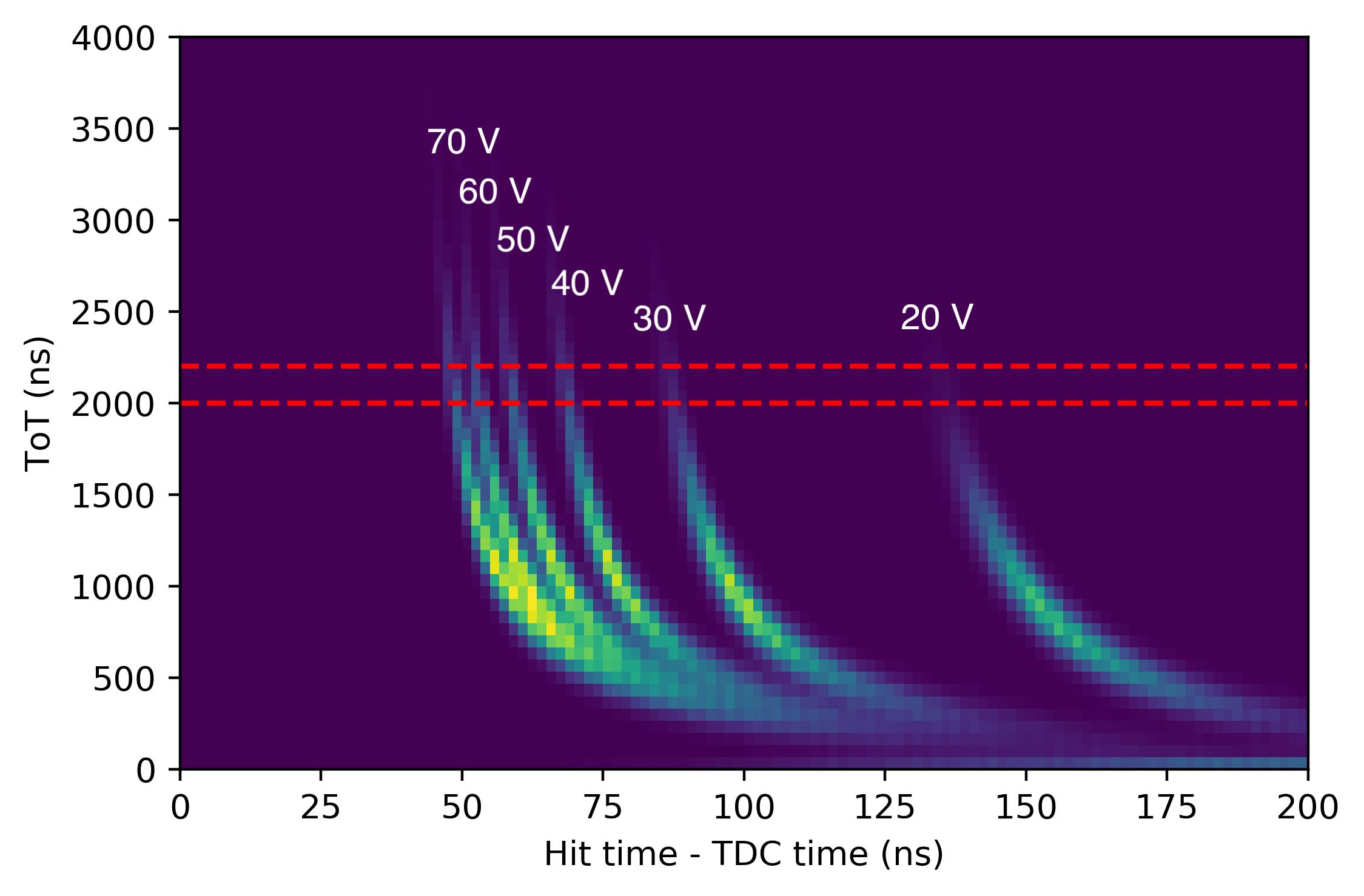}}
		\qquad
		\adjustbox{valign=c}{\includegraphics[width=.47\textwidth]{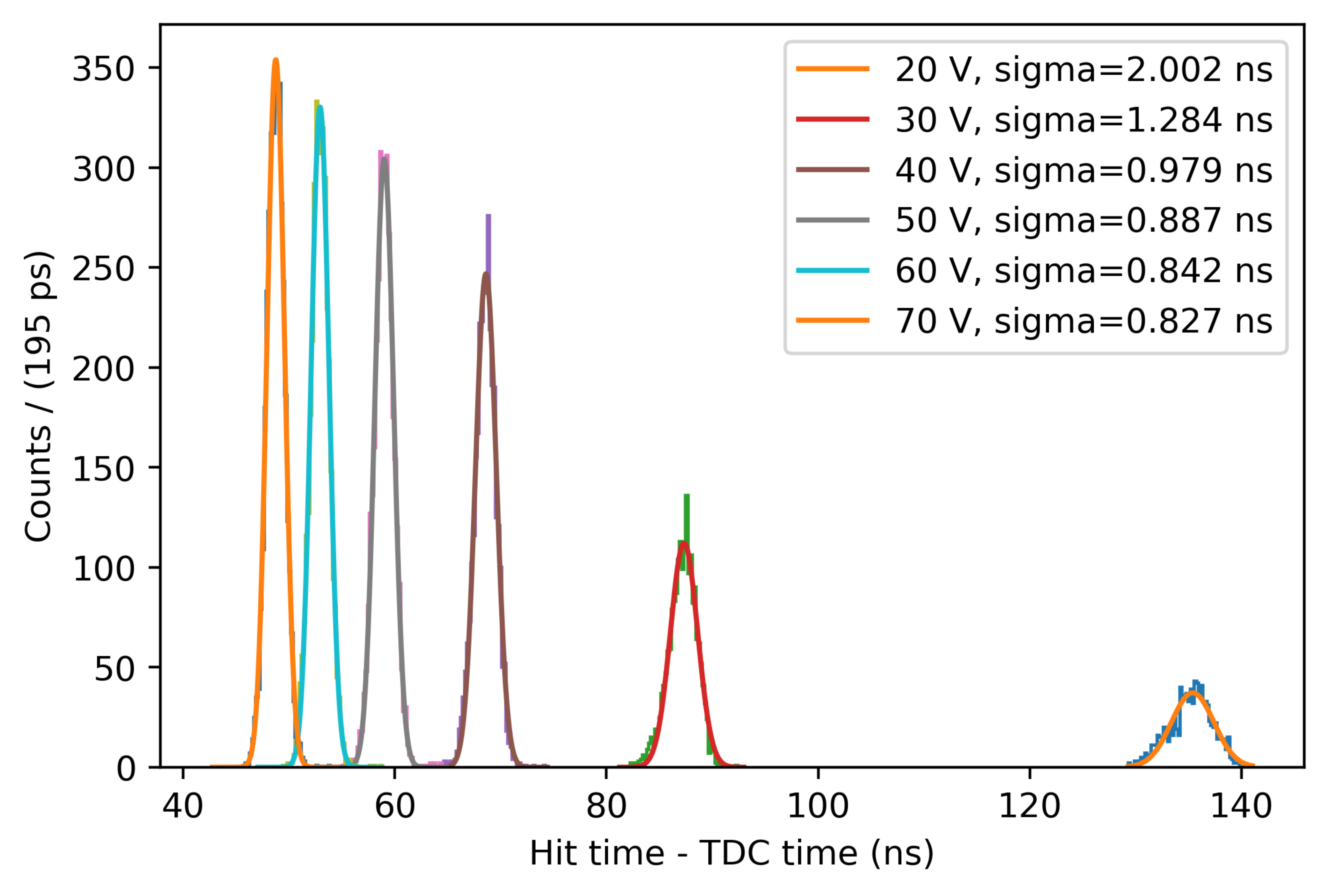}}
		\caption{\label{fig:bias_1} Left: Distribution of ToT versus time difference for multiple values of the bias voltage for the intensified configuration. Right: Distribution of time difference for multiple values of the bias voltage.}
	\end{figure}

Photons emitted by the P47 scintillator in the intensifier are converted to photoelectrons near the sensor surface. Charge collection from the surface depends on the depletion of silicon in this area and, therefore, depends on the applied bias voltage. We measured the dependence of the collection time behaviour on the bias voltage by acquiring datasets for six different voltages: 20, 30, 40, 50, 60, and 70~V in the configuration with an intensifier. Figure \ref{fig:bias_1} shows the ToT versus time difference distributions and individual time difference distributions for all values of the bias voltage. We see that the measured time difference decreases from over 100~ns values for a bias of 20~V when the sensor should start to be depleted to the values around 50~ns for a maximum bias voltage of 70~V.
The latter delay includes the drift time through 300 micron of depleted silicon and additional time offset between the signals. In all cases the timewalk effect is clearly visible. The time difference distributions in the right part of Figure \ref{fig:bias_1} were selected with ToT in the 2000 - 2200~ns range indicated with red lines in the left part of the figure. 

\begin{figure}[ht]
		\centering 
		\adjustbox{valign=c}{\includegraphics[width=.45\textwidth]{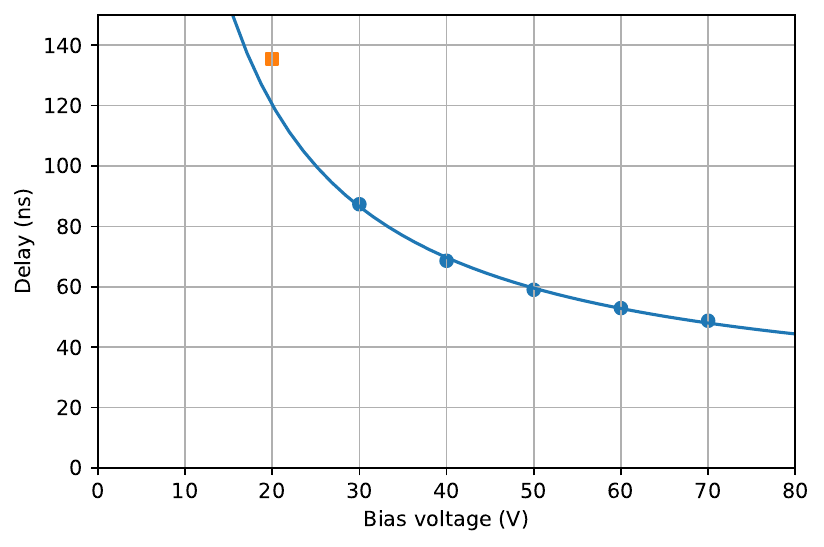}}
		\qquad
		\adjustbox{valign=c}{\includegraphics[width=.45\textwidth]{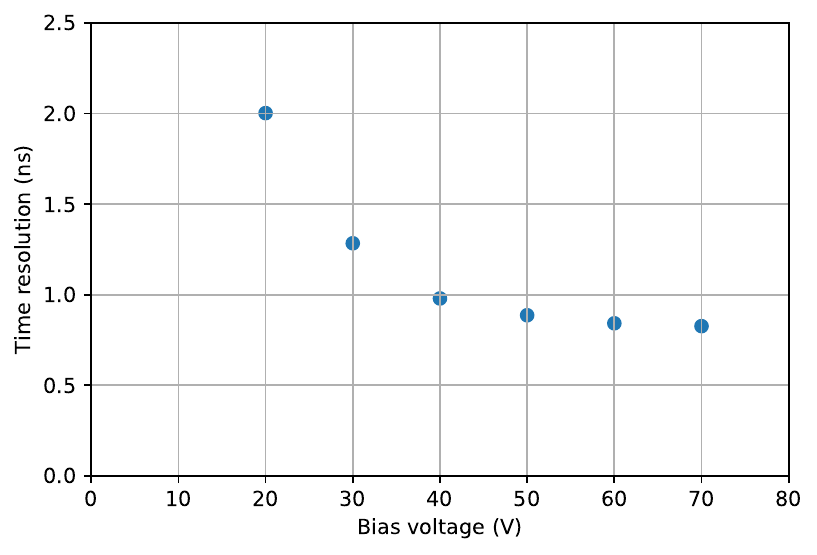}}
		\caption{\label{fig:ToT_scan} Left: Time delay between the laser synchronization signal and ToA of one of the hit Timepix4 pixels as a function of bias voltage. The pixel is activated with a flash of light from the laser with TOT selection in a 200~ns slice. The delay values are fit with a theoretical curve, see the text. The point at 20~V was not used in the fit. 
        Right: Measured time resolution with TOT selection in a 200~ns slice as a function of bias voltage. The 2000-2200~ns ToT slice is indicated with red lines in Figure \ref{fig:bias_1}.}
	\end{figure}

The left and right panels of Figure~\ref{fig:ToT_scan} present, respectively, the time delay between the laser synchronization signal and the ToA of a single Timepix4 pixel activated by a laser flash, and the corresponding time resolution as a function of bias voltage, both obtained with a ToT selection in a 200~ns window from 2000 to 2200~ns.
Assuming the variation in delay with applied voltage~$V$ is entirely caused by the changing drift velocity of the holes through the sensor, we can describe the dependence as
\begin{equation}
    \Delta t = \left( \frac{d^2}{\mu_h} \right) \frac{1}{V} + t_0
    \label{eqdrift}
\end{equation}
with $d$ the sensor thickness, $t_0$ the constant offset due to other effects and $\mu_h$ the hole mobility. 
The dependance of equation~\ref{eqdrift} is only valid for a fully depleted sensor in the low-field region. Based on the sensor properties, full depletion is expected around 27~V. The sensor thus will not be fully depleted for the lowest applied bias of 20~V. In this case, the charge moves through the undepleted region by diffusion rather than drift, which is a slower process and thus introduces additional delays. A precise calculation of this requires detailed knowledge of the sensor physics and is outside of the scope of this work. We therefore omit this measurement from the following analysis. 

The function from equation~\ref{eqdrift} is fitted to the delay in the left part of Figure~\ref{fig:ToT_scan}, yielding a mobility of (4.4 $\pm$ 0.2)$\times$10$^{2}$~cm$^2$V$^{-1}$s$^{-1}$, which is in good agreement with literature values at temperatures slightly above room temperature \cite{Dorkel1981} and thus confirms that the measured delay can be explained by the drift of holes in depleted silicon. The value of the delay for 20~V is approximately 15~ns more than the expected value based on the mobility alone.

The time resolution for the ToT selections, as shown in the right part of Figure~\ref{fig:ToT_scan}, reaches 0.83~ns at 70~V and remains better than 1~ns for a bias voltage as low as 40~V. These results indicate that the ultimate timing performance of the intensified optical camera with Timepix4 readout can be probed at higher bias voltages, which enable prompt charge carrier collection, and at the highest achievable ToT values, which minimize the impact of timewalk corrections and their associated jitter.

The observed time-walk behaviour is primarily governed by the pixel preamplifier peaking time, with an additional contribution from the finite rise and decay time of the P47 scintillator in the intensifier, all convoluted with the discriminator response. While drifting carriers in the depleted silicon introduce some voltage-dependent variation, the overall shape of the time-walk curve remains largely unchanged across the relevant bias range, indicating that weighting-field effects and carrier mobilities play a secondary role. In configurations using n-on-p sensors with electron collection, the intrinsic jitter could in principle be reduced for large signals; however, in the intensified optical scheme the scintillator emission time is the dominant limitation, so only a modest improvement would be expected. Properties of Timepix4 front-end electronics and its implication on time measurement are discussed in detail in \cite{Heijhoff2022,Ballabriga2023,Riegler2017}. 
         
	\section{Conclusions}
	
Timepix3 and Timepix4 are both hybrid-pixel, event-driven readout chips that, when coupled to an optical sensor and image intensifier, enable single-photon, time-resolved optical imaging. Timepix3 provides a 256×256 array (55 µm pitch), 1.56 ns time-of-arrival (ToA), and per-pixel time-over-threshold (ToT). Timepix4 enlarges the matrix to 512×448 at the same pitch (~3.5× area), improves ToA binning to 195 ps, and sustains considerably higher hit rates. Both output ToT, but Timepix4’s upgraded front-end yields more accurate amplitude estimates. Overall, Timepix4 offers superior timing, throughput, and scalability for high-flux, sub-nanosecond optical experiments.

Table \ref{tab:comparison} presents a comparison of timing resolutions (RMS) achieved in this study for different configurations of the Timepix4 camera. The resolution deteriorates after the addition of the intensifier, mostly due to the appearance of a scintillator as an intermediate step in the amplification chain. The P47 scintillator typically used in fast image intensifiers has a rise time of about 7~ns and, therefore, can limit the time resolution in situations with insufficient gain. Faster scintillators are becoming available \cite{Winter2014, Zapadlk2022}, which can be used in intensifiers and also for similar MCP-based amplification schemes employed for ion and electron imaging. Another opportunity to increase the signal amplitude with the aim of improving time resolution is to employ optical sensors based on low-gain avalanche diodes (LGAD), which typically provide a gain of 5 to 10 \cite{Pellegrini2014, Doblas2023, Svihra2024}. Work is in progress to implement and evaluate these options.


\begin{table}[h]
    \centering
    \caption{Timing resolution (RMS) achieved for the optical Timepix4 camera in different configurations.}
    \vspace{0.2cm}
    \begin{tabularx}{\linewidth}{|X|c|X|}
        \hline
        \textbf{Configuration} & \textbf{Time resolution [ns]} & \textbf{Comments} \\
        \hline
        direct flash: no ToT selection & 0.32 & ToT range 300 ns \\
        direct flash: ToT selection & 0.27 & ToT range 25 ns \\
        intensifier: mild ToT selection + timewalk correction &1.0–1.5& large detection efficiency \\
        intensifier: high ToT selection & 0.6–0.7 & small detection efficiency \\
        \hline
    \end{tabularx}
    \label{tab:comparison}
\end{table}
            
In summary, we presented the first characterization of an optical camera based on the Timepix4 chip coupled to an optical silicon sensor and image intensifier, demonstrating its potential for sub-nanosecond scale, time-resolved single-photon imaging. The system achieved time resolutions down to about 0.3~ns without the intensifier and as low as 0.6~ns with the intensifier in the single-photon regime and selection of signals with high amplitude. For two more inclusive cases of amplitude selections covering, respectively, 43\% and 83\% of events, the resolutions with the intensifier are equal to 1.0~ns and to 1.5~ns, which is still a factor of two to three better than the resolution achieved in the past with Timepix3 cameras \cite{Zhao2017coin, Nomerotski2023}. These measurements validate the capability of the Timepix4-based optical cameras to perform fast, high-resolution optical time-stamping. The observed performance marks a significant improvement over previous Timepix3-based systems, particularly in terms of timing precision, pixel count, and readout rate.

Our results also highlight key factors influencing timing performance, including the sensor bias voltage, intensifier scintillator properties, and timewalk effects. After applying a timewalk correction and optimizing selection criteria, we demonstrated sub-nanosecond timing with intensified readout - a performance level suitable for single-photon imaging in quantum optics, ultrafast imaging, and time-correlated photon counting experiments. With its high throughput and improved timing resolution, the Timepix4 camera is well positioned as a scalable platform for next-generation time-resolved optical measurements across a broad range of scientific domains.
	
	\acknowledgments 
	
	This research was supported by the Czech Science Foundation (GACR) under Project No. 25-15534M and Czech Ministry of Education, Youth and Sports Project No. LM2023040 CERN-CZ. This work was supported by the Dutch National Growth Fund (NGF), as part of the Quantum Delta NL Programme. We thank Lou-Ann Pestana De Sousa for help with preparation of the manuscript. 
	
	\bibliography{references} 
	\bibliographystyle{JHEP} 
	
\end{document}